\input amssym

\input epsf
\newfam\scrfam
\batchmode\font\tenscr=rsfs10 \errorstopmode
\ifx\tenscr\nullfont
        \message{rsfs script font not available. Replacing with calligraphic.}
        
\else   
        \font\sevenscr=rsfs7
        \font\fivescr=rsfs5
        \skewchar\tenscr='177 \skewchar\sevenscr='177 \skewchar\fivescr='177
        \textfont\scrfam=\tenscr \scriptfont\scrfam=\sevenscr
        \scriptscriptfont\scrfam=\fivescr

\fi
\catcode`\@=11
\newfam\frakfam
\batchmode\font\tenfrak=eufm10 \errorstopmode
\ifx\tenfrak\nullfont
        \message{eufm font not available. Replacing with italic.}
        
\else
	
	\font\sevenfrak=eufm7 \font\fivefrak=eufm5
	\textfont\frakfam=\tenfrak
	\scriptfont\frakfam=\sevenfrak \scriptscriptfont\frakfam=\fivefrak
	
\fi
\catcode`\@=\active
\newfam\msbfam
\batchmode\font\twelvemsb=msbm10 scaled\magstep1 \errorstopmode
\ifx\twelvemsb\nullfont\def\Bbb{\bf}

	\message{Blackboard bold not available. Replacing with boldface.}
\else   \catcode`\@=11
        \font\tenmsb=msbm10 \font\sevenmsb=msbm7 \font\fivemsb=msbm5
        \textfont\msbfam=\tenmsb
        \scriptfont\msbfam=\sevenmsb \scriptscriptfont\msbfam=\fivemsb
        \def\Bbb{\relax\expandafter\Bbb@}
        \def\Bbb@#1{{\Bbb@@{#1}}}
        \def\Bbb@@#1{\fam\msbfam\relax#1}
        \catcode`\@=\active

\fi
        \font\fivemi=cmmi5
        \font\sixmi=cmmi6
        \font\eightrm=cmr8              \def\xrm{\eightrm}
        \font\eightbf=cmbx8             \def\xbf{\eightbf}
        \font\eightit=cmti10 at 8pt     \def\xit{\eightit}
                
        \font\eighttt=cmtt8             
        \font\eightcp=cmcsc8
        \font\eighti=cmmi8              \def\xold{\eighti}
        \font\eightmi=cmmi8
        \font\eightib=cmmib8             \def\xbold{\eightib}
        \font\teni=cmmi10               \def\old{\teni}
        \font\tencp=cmcsc10
        
        \font\twelverm=cmr12
        
        \font\twelvecp=cmcsc10 scaled\magstep1
        \font\fourteencp=cmcsc10 scaled\magstep2
        \font\sixrm=cmr6
        \font\fiverm=cmr5
        
        \font\eightmath=cmmi8
	
        \font\eightsy=cmsy8
        \font\sixsy=cmsy6
        \font\eightsl=cmsl8
        \font\sixbf=cmbx6

	 at10pt	
	 at12pt
	 at14pt
	 at16pt
	
	\font\fourteenmath=cmmi14

\def\noblackbox{\overfullrule=0pt}
\noblackbox

\def\eightpoint{
\def\rm{\fam0\eightrm}
\textfont0=\eightrm \scriptfont0=\sixrm \scriptscriptfont0=\fiverm
\textfont1=\eightmi  \scriptfont1=\sixmi  \scriptscriptfont1=\fivemi
\textfont2=\eightsy \scriptfont2=\sixsy \scriptscriptfont2=\fivesy
\textfont3=\tenex   \scriptfont3=\tenex \scriptscriptfont3=\tenex
\textfont\itfam=\eightit \def\it{\fam\itfam\eightit}
\textfont\slfam=\eightsl \def\sl{\fam\slfam\eightsl}
\textfont\ttfam=\eighttt \def\tt{\fam\ttfam\eighttt}
\textfont\bffam=\eightbf \scriptfont\bffam=\sixbf 
                         \scriptscriptfont\bffam=\fivebf
                         \def\bf{\fam\bffam\eightbf}
\normalbaselineskip=10pt}

\newtoks\headtext
\headline={\ifnum\pageno=1\hfill\else
	\ifodd\pageno{\eightcp\the\headtext}{ }\dotfill{ }{\old\folio}
	\else{\old\folio}{ }\dotfill{ }{\eightcp\the\headtext}\fi
	\fi}
\def\makeheadline{\vbox to 0pt{\vss\noindent\the\headline\break
\hbox to\hsize{\hfill}}
        \vskip2\baselineskip}
\newcount\infootnote
\infootnote=0
\def\foot#1#2{\infootnote=1
\footnote{${}^{#1}$}{\vtop{\baselineskip=.75\baselineskip
\advance\hsize by
-\parindent{\eightpoint\rm\hskip-\parindent #2}\hfill\vskip\parskip}}\infootnote=0}
\newcount\refcount
\refcount=1
\newwrite\refwrite
\def\oldsize{\ifnum\infootnote=1\xold\else\old\fi}
\def\ref#1#2{
	\def#1{{{\oldsize\the\refcount}}\ifnum\the\refcount=1\immediate\openout\refwrite=\jobname.refs\fi\immediate\write\refwrite{\item{[{\xold\the\refcount}]} 
	#2\hfill\par\vskip-2pt}\xdef#1{{\noexpand\oldsize\the\refcount}}\global\advance\refcount by 1}
	}
\def\refout{\catcode`\@=11
        \xrm\immediate\closeout\refwrite
        \vskip2\baselineskip
        {\noindent\twelvecp References}\hfill\vskip\baselineskip
        \baselineskip=.75\baselineskip
        \input\jobname.refs
        \baselineskip=4\baselineskip \divide\baselineskip by 3
        \catcode`\@=\active\rm}

\def\skipref#1{\hbox to15pt{\phantom{#1}\hfill}\hskip-15pt}

\def\hepth#1{\href{http://xxx.lanl.gov/abs/hep-th/#1}{arXiv:hep-th/{\xold#1}}}

\def\arxiv#1#2{\href{http://arxiv.org/abs/#1.#2}{arXiv:{\xold#1}.{\xold#2}}}
\def\jhep#1#2#3#4{\href{http://jhep.sissa.it/stdsearch?paper=#2\%28#3\%29#4}{J. High Energy Phys. {\xbold #1#2} ({\xold#3}) {\xold#4}}}

\def\ATMP#1#2#3{Adv. Theor. Math. Phys. {\xbold#1} ({\xold#2}) {\xold#3}}

\def\CQG#1#2#3{Class. Quantum Grav. {\xbold#1} ({\xold#2}) {\xold#3}}
\def\FP#1#2#3{Fortschr. Phys. {\xbold#1} ({\xold#2}) {\xold#3}}

\def\JGP#1#2#3{J. Geom. Phys. {\xbold#1} ({\xold#2}) {\xold#3}}
\def\JHEP{\jhep}

\def\LMP#1#2#3{Lett. Math. Phys. {\xbold#1} ({\xold#2}) {\xold#3}}
\def\MPLA#1#2#3{Mod. Phys. Lett. {\xbf A}{\xbold#1} ({\xold#2}) {\xold#3}}

\def\NPB#1#2#3{Nucl. Phys. {\xbf B}{\xbold#1} ({\xold#2}) {\xold#3}}

\def\PLB#1#2#3{Phys. Lett. {\xbf B}{\xbold#1} ({\xold#2}) {\xold#3}}

\def\PTPS#1#2#3{Progr. Theor. Phys. Suppl. {\xbold#1} ({\xold#2}) {\xold#3}}
\newcount\sectioncount
\sectioncount=0
\def\section#1#2{\global\eqcount=0
	\global\subsectioncount=0
        \global\advance\sectioncount by 1
	\ifnum\sectioncount>1
	        \vskip2\baselineskip
	\fi
\line{\twelvecp\the\sectioncount. #2\hfill}
       \vskip.5\baselineskip\noindent
        \xdef#1{{\old\the\sectioncount}}}
\newcount\subsectioncount
\def\subsection#1#2{\global\advance\subsectioncount by 1
\vskip.75\baselineskip\noindent\line{\tencp\the\sectioncount.\the\subsectioncount. #2\hfill}\nobreak\vskip.4\baselineskip\nobreak\noindent\xdef#1{{\old\the\sectioncount}.{\old\the\subsectioncount}}}
\def\immediatesubsection#1#2{\global\advance\subsectioncount by 1
\vskip-\baselineskip\noindent
\line{\tencp\the\sectioncount.\the\subsectioncount. #2\hfill}
	\vskip.5\baselineskip\noindent
	\xdef#1{{\old\the\sectioncount}.{\old\the\subsectioncount}}}
\newcount\appendixcount
\appendixcount=0
\def\appendix#1{\global\eqcount=0
        \global\advance\appendixcount by 1
        \vskip2\baselineskip\noindent
        \ifnum\the\appendixcount=1
        \hbox{\twelvecp Appendix A: #1\hfill}\vskip\baselineskip\noindent\fi
    \ifnum\the\appendixcount=2
        \hbox{\twelvecp Appendix B: #1\hfill}\vskip\baselineskip\noindent\fi
    \ifnum\the\appendixcount=3
        \hbox{\twelvecp Appendix C: #1\hfill}\vskip\baselineskip\noindent\fi}

\newcount\eqcount
\eqcount=0
\def\Eqn#1{\global\advance\eqcount by 1
\ifnum\the\sectioncount=0
	\xdef#1{{\noexpand\oldsize\the\eqcount}}
	\eqno({\oldstyle\the\eqcount})
\else
        \ifnum\the\appendixcount=0
\xdef#1{{\noexpand\oldsize\the\sectioncount}.{\noexpand\oldsize\the\eqcount}}
                \eqno({\oldstyle\the\sectioncount}.{\oldstyle\the\eqcount})\fi
        \ifnum\the\appendixcount=1
	        \xdef#1{{\noexpand\oldstyle A}.{\noexpand\oldstyle\the\eqcount}}
                \eqno({\oldstyle A}.{\oldstyle\the\eqcount})\fi
        \ifnum\the\appendixcount=2
	        \xdef#1{{\noexpand\oldstyle B}.{\noexpand\oldstyle\the\eqcount}}
                \eqno({\oldstyle B}.{\oldstyle\the\eqcount})\fi
        \ifnum\the\appendixcount=3
	        \xdef#1{{\noexpand\oldstyle C}.{\noexpand\oldstyle\the\eqcount}}
                \eqno({\oldstyle C}.{\oldstyle\the\eqcount})\fi
\fi}
\def\eqn{\global\advance\eqcount by 1
\ifnum\the\sectioncount=0
	\eqno({\oldstyle\the\eqcount})
\else
        \ifnum\the\appendixcount=0
                \eqno({\oldstyle\the\sectioncount}.{\oldstyle\the\eqcount})\fi
        \ifnum\the\appendixcount=1
                \eqno({\oldstyle A}.{\oldstyle\the\eqcount})\fi
        \ifnum\the\appendixcount=2
                \eqno({\oldstyle B}.{\oldstyle\the\eqcount})\fi
        \ifnum\the\appendixcount=3
                \eqno({\oldstyle C}.{\oldstyle\the\eqcount})\fi
\fi}
\def\multi{\global\advance\eqcount by 1}
\def\multieq#1#2{\xdef#1{{\old\the\eqcount#2}}
        \eqno{({\oldstyle\the\eqcount#2})}}
\newtoks\url
\def\Href#1#2{\catcode`\#=12\url={#1}\catcode`\#=\active#2}
\def\href#1#2{{#2}}

\parskip=3.5pt plus .3pt minus .3pt
\baselineskip=14pt plus .1pt minus .05pt
\lineskip=.5pt plus .05pt minus .05pt
\lineskiplimit=.5pt
\abovedisplayskip=18pt plus 4pt minus 2pt
\belowdisplayskip=\abovedisplayskip
\hsize=14cm
\vsize=19cm
\hoffset=1.5cm
\voffset=1.8cm
\frenchspacing
\footline={}
\raggedbottom

\newskip\origparindent
\origparindent=\parindent

\def\ss{\scriptstyle}

\def\*{\partial}
\def\punkt{\,\,.}
\def\komma{\,\,,}

\def\={\!=\!}
\def\small#1{{\hbox{$#1$}}}

\def\fraction#1{\small{1\over#1}}
\def\fr{\fraction}
\def\Fraction#1#2{\small{#1\over#2}}
\def\Fr{\Fraction}

\def\eg{{\it e.g.}}

\def\ie{{\it i.e.}}

\def\nlni{\hfill\break}

\def\a{\alpha}
\def\b{\beta}
\def\d{\delta}

\def\g{\gamma}
\def\l{\lambda}



\def\lb{\bar\l}

\ref\ABC{Y. Aisaka and M. Cederwall, unpublished.}

\ref\BerkovitsNonMinimal{N. Berkovits,
{\xit ``Pure spinor formalism as an N=2 topological string''},
\jhep{05}{10}{2005}{089} [\hepth{0509120}].}

\ref\BerkovitsVanishing{N. Berkovits, {\xit ``Multiloop amplitudes and
vanishing theorems using the pure spinor formalism for the
superstring''}, \jhep{04}{09}{2004}{047} [\hepth{0406055}].}

\ref\BerkovitsNekrasovMultiloop{N. Berkovits and N. Nekrasov, {\xit
    ``Multiloop superstring amplitudes from non-minimal pure spinor
    formalism''}, \jhep{06}{12}{2006}{029} [\hepth{0609012}].}

\ref\MovshevEleven{M.V. Movshev,	
{\xit ``Geometry of a desingularization of eleven-dimensional
gravitational spinors''}, \hfill\break\arxiv{1105}{0127}.}

\ref\PureSGI{M. Cederwall, {\xit ``Towards a manifestly supersymmetric
    action for D=11 supergravity''}, \jhep{10}{01}{2010}{117}
    [\arxiv{0912}{1814}].}  

\ref\PureSGII{M. Cederwall, 
{\xit ``D=11 supergravity with manifest supersymmetry''},
    \MPLA{25}{2010}{3201} [\arxiv{1001}{0112}].}

\ref\GrassiGuttenberg{P.A. Grassi and S. Guttenberg, 
{\xit ``On projections to the pure spinor space''},
\arxiv{1109}{2848}.}

\ref\GriffithsHarris{P. Griffiths and J. Harris, {\xit ``Principles of
algebraic geometry''}, Wiley \&\ Sons (1978).}

\ref\CederwallNilssonTsimpisI{M. Cederwall, B.E.W. Nilsson and D. Tsimpis,
{\xit ``The structure of maximally supersymmetric super-Yang--Mills
theory --- constraining higher order corrections''},
\jhep{01}{06}{2001}{034} 
[\hepth{0102009}].}

\ref\CederwallNilssonTsimpisII{M. Cederwall, B.E.W. Nilsson and D. Tsimpis,
{\xit ``D=10 super-Yang--Mills at $\ss O(\a'^2)$''},
\JHEP{01}{07}{2001}{042} [\hepth{0104236}].}

\ref\BerkovitsParticle{N. Berkovits, {\xit ``Covariant quantization of
the superparticle using pure spinors''}, \jhep{01}{09}{2001}{016}
[\hepth{0105050}].}

\ref\SpinorialCohomology{M. Cederwall, B.E.W. Nilsson and D. Tsimpis,
{\xit ``Spinorial cohomology and maximally supersymmetric theories''},
\jhep{02}{02}{2002}{009} [\hepth{0110069}];
M. Cederwall, {\xit ``Superspace methods in string theory, supergravity and gauge theory''}, Lectures at the XXXVII Winter School in Theoretical Physics ``New Developments in Fundamental Interactions Theories'',  Karpacz, Poland,  Feb. 6-15, 2001, \hepth{0105176}.}

\ref\Movshev{M. Movshev and A. Schwarz, {\xit ``On maximally
supersymmetric Yang--Mills theories''}, \NPB{681}{2004}{324}
[\hepth{0311132}].}

\ref\BerkovitsI{N. Berkovits,
{\xit ``Super-Poincar\'e covariant quantization of the superstring''},
\jhep{00}{04}{2000}{018} [\hepth{0001035}].}

\ref\CederwallNilssonSix{M. Cederwall and B.E.W. Nilsson, {\xit ``Pure
spinors and D=6 super-Yang--Mills''}, \arxiv{0801}{1428}.}

\ref\CGNN{M. Cederwall, U. Gran, M. Nielsen and B.E.W. Nilsson,
{\xit ``Manifestly supersymmetric M-theory''},
\JHEP{00}{10}{2000}{041} [\hepth{0007035}];
{\xit ``Generalised 11-dimensional supergravity''}, \hepth{0010042}.
}

\ref\CGNT{M. Cederwall, U. Gran, B.E.W. Nilsson and D. Tsimpis,
{\xit ``Supersymmetric corrections to eleven-dimen\-sional supergravity''},
\jhep{05}{05}{2005}{052} [\hepth{0409107}].}

\ref\NilssonPure{B.E.W.~Nilsson,
{\xit ``Pure spinors as auxiliary fields in the ten-dimensional
supersymmetric Yang--Mills theory''},
\CQG3{1986}{{\xrm L}41}.}

\ref\HowePureI{P.S. Howe, {\xit ``Pure spinor lines in superspace and
ten-dimensional supersymmetric theories''}, \PLB{258}{1991}{141}.}

\ref\HowePureII{P.S. Howe, {\xit ``Pure spinors, function superspaces
and supergravity theories in ten and eleven dimensions''},
\PLB{273}{1991}{90}.} 



\ref\CederwallThreeConf{M. Cederwall, {\xit ``N=8 superfield formulation of
the Bagger--Lambert--Gustavsson model''}, \jhep{08}{09}{2008}{116}
[\arxiv{0808}{3242}]; {\xit ``Superfield actions for N=8 
and N=6 conformal theories in three dimensions''},
\jhep{08}{10}{2008}{70}
[\arxiv{0808}{3242}]; {\xit ``Pure spinor superfields,
with application to D=3 conformal models''}, \arxiv{0906}{5490}.}

\ref\BerkovitsICTP{N. Berkovits, {\xit ``ICTP lectures on covariant
quantization of the superstring''}, proceedings of the ICTP Spring
School on Superstrings and Related Matters, Trieste, Italy, 2002
[\hepth{0209059}.]} 

\ref\BatalinVilkovisky{I.A. Batalin and G.I. Vilkovisky, {\xit ``Gauge
algebra and quantization''}, \PLB{102}{1981}{27}.}

\ref\FusterBVReview{A. Fuster, M. Henneaux and A. Maas, {\xit
``BRST-antifield quantization: a short review''},\nlni\hepth{0506098}.}

\ref\BerkovitsMembrane{N. Berkovits,
	{\xit ``Towards covariant quantization of the supermembrane''},
	\JHEP{02}{09}{2002}{051} [\hepth{0201151}].}

\ref\BerkovitsNekrasovCharacter{N. Berkovits and N. Nekrasov, {\xit
    ``The character of pure spinors''}, \LMP{74}{2005}{75}
  \nlni[\hepth{0503075}].}

\ref\AnguelovaGrassiVanhove{L. Anguelova, P.A. Grassi and P. Vanhove,
  {\xit ``Covariant one-loop amplitudes in D=11''},
  \NPB{702}{2004}{269} [\hepth{0408171}].}

\ref\GrassiVanhove{P.A. Grassi and P. Vanhove, {\xit ``Topological M
    theory from pure spinor formalism''}, \ATMP{9}{2005}{285}
  [\hepth{0411167}].} 

\ref\BerkovitsNekrasovMultiloop{N. Berkovits and N. Nekrasov, {\xit
    ``Multiloop superstring amplitudes from non-minimal pure spinor
    formalism''}, \jhep{06}{12}{2006}{029} [\hepth{0609012}].}

\ref\BerkovitsII{N. Berkovits and B.C. Valillo, 
{\xit ``Consistency of super-Poincar\'e covariant superstring tree
amplitudes''}, \jhep{00}{07}{2000}{015} [\hepth{0004171}].}

\ref\BjornssonGreen{J. Bj\"ornsson and M.B. Green, {\xit ``5 loops in
25/4 dimensions''}, \jhep{10}{08}{2010}{132} [\arxiv{1004}{2692}].}

\ref\BjornssonMultiLoop{J. Bj\"ornsson, {\xit ``Multi-loop amplitudes
in maximally supersymmetric pure spinor field
theory''}, \jhep{11}{01}{2011}{002} [\arxiv{1009}{5906}].}

\ref\BerkovitsCherkis{N. Berkovits and S.A. Cherkis, {\xit
``Higher-dimensional twistor transforms using pure spinors''}, 
\jhep{04}{12}{2004}{049} [\hepth{0409243}].}

\ref\CederwallKarlssonBI{M. Cederwall and A. Karlsson, {\xit ``Pure
spinor superfields and Born--Infeld theory''},
\jhep{11}{11}{2011}{134} [\arxiv{1109}{0809}].}

\ref\GomezOneLoop{H. Gomez, {\xit ``One-loop superstring amplitude
from integrals on pure spinors space''}, \jhep{09}{12}{2009}{034}
[\arxiv{0910}{3405}].} 

\ref\NekrasovBetaGamma{N. Nekrasov, {\xit ``Lectures on curved
beta-gamma systems, pure spinors, and anomalies''}, \hepth{0511008}.}

\ref\AisakaArroyoBerkovitsNekrasov{Y. Aisaka, E.A. Arroyo,
N. Berkovits and N. Nekrasov, {\xit ``Pure spinor partition function
and the massive superstring spectrum''}, \jhep{08}{08}{2008}{050}
[\arxiv{0806}{0584}].} 

\ref\AisakaArroyo{Y. Aisaka and E.A. Arroyo, {\xit ``Hilbert space of
curved $\ss\beta\gamma$ systems on quadric cones''},
\jhep{08}{08}{2008}{052} [\arxiv{0806}{0586}].}

\ref\AisakaOperatorSpace{Y. Aisaka, {\xit ``Operator space of pure
spinors''}, \PTPS{188}{2011}{227}.}

\ref\CederwallPureSpinorSpace{M. Cederwall, {\xit ``The geometry of
pure spinor space''}, \jhep{12}{01}{2012}{150} 
\hfill\break[\arxiv{1111}{1932}].}

\ref\CederwallOperators{M. Cederwall, {\xit ``Operators on pure spinor
space''}, AIP Conf. Proc. {\xbold1243} ({\xold2010}) {\xold51}.}

\ref\PureSpinorOverview{M. Cederwall, {\xit ``Pure spinor superfields
--- an overview''}, Springer Proc. Phys. {\xbf153} ({\xrm2013}) {\xrm61} 
[\arxiv{1307}{1762}].}

\ref\CederwallPureSpinorCQG{M. Cederwall, {\xit ``Pure spinors in classical and quantum supergravity''}, \arxiv{2210}{06141}, to appear in 
{\xit ``Handbook of quantum gravity''}, eds. C. Bambi, L. Modesto and I. Shapiro.}

\ref\CederwallJonssonPalmkvistSaberi{M. Cederwall, S. Jonsson, J. Palmkvist and I. Saberi,
{\xit ``Supersymmetry and Koszul duality''}, to appear.}

\ref\CederwallPureSpinorSpace{M. Cederwall, {\xit ``The geometry of
pure spinor space''}, \jhep{12}{01}{2012}{150}  
\hfill\break[\arxiv{1111}{1932}].}


\ref\OdaToninB{I. Oda and M. Tonin, {\xit ``Y-formalism and b ghost in the non-minimal pure spinor formalism of superstrings''},
\NPB{779}{2007}{63} [\arxiv{0704}{1219}].}

\ref\BerkovitsMazzucato{N. Berkovits and L. Mazzucato, {\xit ``Taming the b antighost with Ramond-Ramond flux''},
\jhep{10}{11}{2010}{019} [\arxiv{1004}{5140}].}

\ref\ChandiaB{O. Chandia, {\xit ``The b ghost of the pure spinor formalism is nilpotent''}, 
\PLB{695}{2011}{312} [\arxiv{1008}{1778}].}

\ref\LipinskiBNilpot{R. Lipinski Jusinskas, {\xit ``Nilpotency of the b ghost in the non-minimal pure spinor formalism''}, 
\jhep{13}{05}{2013}{048} [\arxiv{1303}{3966}].}

\ref\BerkovitsBTwist{N. Berkovits, {\xit ``Dynamical twisting and the b ghost in the pure spinor formalism''},
\jhep{13}{06}{2013}{091} [\arxiv{1305}{0693}].}

\ref\LipinskiB{R. Lipinski Jusinskas, {\xit ``Notes on the pure spinor b ghost''}, \jhep{13}{07}{2013}{142}
[\arxiv{1306}{1963}].}

\ref\BakhmatovBerkovitsB{I. Bakhmatov and N. Berkovits, {\xit ``Pure spinor b-ghost in a super--Maxwell background''},
\jhep{13}{11}{2013}{214} [\arxiv{1310}{3379}].}

\ref\LipinskiThesis{R. Lipinski Jusinskas, {\xit ``Exploring the properties of the pure spinor b ghost''}, PhD thesis, ITP S\~ao Paulo (2014).} 

\ref\FleuryB{T. Fleury, {\xit ``On the pure spinor heterotic superstring b ghost''},
\jhep{16}{03}{2016}{200} [\arxiv{1512}{00807}].}

\ref\BerkovitsGuillenB{N. Berkovits and M. Guillen, {\xit ``Simplified D=11 pure spinor b ghost''}, 
\jhep{17}{07}{2017}{115} [\arxiv{1703}{05116}].}

\ref\ChandiaValliloB{O. Chandia and B.C. Vallilo, {\xit ``Relating the 
b ghost and the vertex operators of the pure spinor superstring''},\jhep{21}{03}{2021}{165} 
[\arxiv{2101}{01129}].}

\ref\PureSGI{M. Cederwall, {\xit ``Towards a manifestly supersymmetric
    action for D=11 supergravity''}, \jhep{10}{01}{2010}{117}
    [\arxiv{0912}{1814}].}  

\ref\PureSGII{M. Cederwall, 
{\xit ``D=11 supergravity with manifest supersymmetry''},
    \MPLA{25}{2010}{3201} [\arxiv{1001}{0112}].}

\ref\GuillenTamingElevenB{M. Guillen, {\xit `Taming the 11D pure spinor b-ghost''},
\arxiv{2212}{13653}.}

\ref\BenShaharGuillen{M. Ben-Shahar and M. Guillen, {\xit ``10D super-Yang--Mills scattering amplitudes from its pure spinor action''},
\jhep{21}{12}{2021}{014} [\arxiv{2108}{11708}].}

\ref\BorstenKinematic{L. Borsten, B. Jur\v co, H. Kim, T. Macrelli and C. S\"amann,
{\xit ``Kinematic Lie algebras from twistor spaces''}, \arxiv{2211}{13261}.}

\ref\EagerHahnerSaberiWilliams{R. Eager, F. Hahner, I. Saberi and B. Williams, {\xit ``Perspectives on the pure spinor superfield formalism''},
\JGP{180}{2022}{104626} [\arxiv{2111}{01162}].}

\ref\CederwallBLG{M. Cederwall, {\xit ``N=8 superfield formulation of
the Bagger--Lambert--Gustavsson model''}, \jhep{08}{09}{2008}{116}
[\arxiv{0808}{3242}].}

\ref\CederwallABJM{M. Cederwall, {\xit ``Superfield actions for N=8 
and N=6 conformal theories in three dimensions''},
\jhep{08}{10}{2008}{70}
[\arxiv{0809}{0318}].}

\ref\CederwallReformulation{M. Cederwall, {\xit ``An off-shell superspace
reformulation of D=4, N=4 super-Yang--Mills theory''},
\FP{66}{2018}{1700082} [\arxiv{1707}{00554}].}

\ref\CederwallDSix{M. Cederwall, {\xit ``Pure spinor superspace action
for D=6, N=1 super-Yang--Mills theory''}, \jhep{18}{05}{2018}{115} [\arxiv{1712}{02284}].}

\ref\CederwallExotic{M. Cederwall {\xit ``Superspace formulation of
exotic supergravities in six dimensions''}, \jhep{21}{03}{2021}{56} [\arxiv{2012}{02719}].}

\ref\CederwallSLFive{M. Cederwall, {\xit ``SL(5) supersymmetry''}, \FP{69}{2021}{2100116} [\arxiv{2107}{09037}].}


\def\textfrac#1#2{\raise .45ex\hbox{\the\scriptfont0 #1}\nobreak\hskip-1pt/\hskip-1pt\hbox{\the\scriptfont0 #2}}

\def\db{\bar\*}

\def\tw{\tilde w}

\def\Sym{\hbox{Sym}}

%
\line{
\epsfysize=18mm
\epsffile{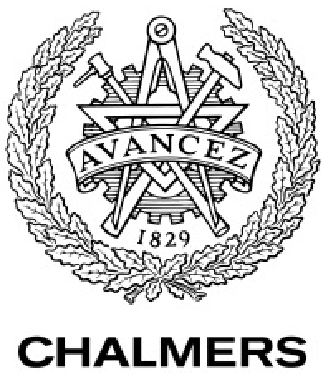}
\hfill}
\vskip-16mm

\line{\hfill}
\line{\hfill Gothenburg preprint}
\line{\hfill December, {\old2022}}
\line{\hrulefill}


\headtext={Cederwall: 
``A minimal {\eightmath b} ghost''}

\vfill

\centerline{\fourteencp
A minimal {\fourteenmath b} ghost}


%

\vfill

\centerline{\twelverm Martin Cederwall}

\vfill
\vskip-1cm

\centerline{\it Department of Physics, Chalmers Univ. of Technology, SE-412 96 Gothenburg, Sweden}

\vskip\parskip
\centerline{and}
\vskip\parskip

\centerline{\it NORDITA, Hannes Alfv\'ens v\"ag 12, SE-106 91 Stockholm, Sweden}

\vfill

{\narrower\noindent \underbar{Abstract:}
The $b$ ghost, or $b$ operator, used for fixing Siegel gauge in the pure spinor superfield formalism, is a composite operator of negative ghost number, satisfying $\{q,b\}=\square$, where $q$ is the pure spinor differential (BRST operator).
It is traditionally constructed using non-minimal variables. 
However, since all cohomology has minimal representatives, it seems likely that there should be versions of physically meaningful operators, also with negative ghost number, using only minimal variables.
The purpose of this letter is to demonstrate that this statement holds by providing a concrete construction in $D=10$ super-Yang--Mills theory, and to argue that it is a general feature in the pure spinor superfield formalism.
\smallskip}
\vfill

\font\xxtt=cmtt6

\vtop{\baselineskip=.6\baselineskip\xxtt
\line{\hrulefill}
\catcode`\@=11
\line{email: martin.cederwall@chalmers.se\hfill}
\catcode`\@=\active
}

\eject


\section\IntroSec{Introduction}The pure spinor superfield formalism (see \eg\ refs. 
[\PureSpinorOverview,\CederwallPureSpinorCQG,\EagerHahnerSaberiWilliams]) is a general method to formulate supersymmetric field theory, which for certain theories allows for an off-shell superfield formulation, even with maximal supersymmetry.
Due to the absence of world-line reparametrisation invariance in the corresponding particle models,
the $b$ ghost is absent as a primitive operator, and the analogous operator (with the same name) needs to be constructed as a composite operator.
Even the classical formulation of interactions sometimes demands negative ghost number operators. This applies to $D=11$ supergravity
[\PureSGI,\PureSGII], and to higher derivative terms like Born--Infeld theory [\CederwallKarlssonBI]. Such operators typically has a physical meaning in terms of mapping between different cohomology classes.
They are traditionally constructed using non-minimal variables, and have quite complicated expressions.
The purpose of this letter is to demonstrate how this may be done in the minimal picture. The main focus will be on $D=10$ super-Yang--Mills theory.

\section\NonMinBSec{The non-minimal $b$ operator}The $b$ ghost, or $b$ operator, used for fixing Siegel gauge in the pure spinor superfield formalism, 
is a composite operator of negative ghost number, satisfying $\{Q,b\}=\square$, where 
$Q$ is some version of the pure spinor differential (BRST operator).
In minimal variables, consisting of $x^a$, $\theta^\alpha$ and $\lambda^\alpha$ with $(\lambda\gamma^a\lambda)=0$, the differential is
$$
q=\lambda^\alpha D_\alpha\;.\eqn
$$ 

The $b$ operator is traditionally constructed using non-minimal variables [\BerkovitsNonMinimal]. 
Then, also $\bar\lambda_\alpha$ and $r_\alpha=d\bar\lambda_\alpha$ are included. This is natural from many points of view, including integration
[\BerkovitsNonMinimal,\CederwallPureSpinorSpace].
The differential is modified to $Q=\lambda^\alpha D_\alpha+\bar\*$, where $\bar\*=d\bar\lambda_\alpha{\*\over\*\bar\lambda_\alpha}$ is the Dolbeault operator on pure spinor space.

However, since all cohomology has minimal representatives, it seems likely that there should be versions of physically meaningful operators, also with negative ghost number, using only minimal variables, \ie, there should exist an operator $b'$ acting on holomorphic functions of $\lambda$, such that
$\{q,b'\}=\square$.
The purpose of this letter is to demonstrate that this statement holds by providing a concrete construction in $D=10$ super-Yang--Mills theory, and arguing that it is a general feature in the pure spinor superfield formalism.

The construction and examination of the $b$ operator (mainly for $D=10$ super-Yang--Mills theory and $D=11$ supergravity)
has been the subject of many papers
[\OdaToninB\skipref\BerkovitsMazzucato\skipref\ChandiaB\skipref\LipinskiBNilpot\skipref\BerkovitsBTwist\skipref\LipinskiB\skipref\BakhmatovBerkovitsB\skipref\LipinskiThesis\skipref\FleuryB\skipref\BerkovitsGuillenB\skipref\ChandiaValliloB-\GuillenTamingElevenB],
mainly because this and similar issues is where the pure spinor superfield formalism becomes more complicated.
In the present letter, we will
work in $D=10$ super-Yang--Mills theory, although analogous minimal operators should always exist.

\noindent The $b$ operator is given
in non-minimal pure spinor variables as the cochain 
[\BerkovitsNonMinimal]
$$
\eqalign{
b&=b_0+b_1+b_2+b_3\cr
&=-\fr2(\l\lb)^{-1}(\lb\g^aD)\*_a
+\fr{16}(\l\lb)^{-2}(\lb\g^{abc}d\lb)\left[N_{ab}\*_c
                         -\fr{24}(D\g_{abc}D)\right]\cr
&+\fr{64}(\l\lb)^{-3}(d\lb\g^{abc}d\lb)(\lb\g_aD)N_{bc}
-\fr{1024}(\l\lb)^{-4}(\lb\g^{ab}{}_id\lb)(d\lb\g^{cdi}d\lb)N_{ab}N_{cd}
\punkt\cr}\eqn
$$
For useful rewritings, see in particular ref. [\BenShaharGuillen].
We use conventions where the torsion is
$T_{\a\b}{}^a=2\g^a_{\a\b}$, \ie, $\{D_\a,D_\b\}=-2\g^a_{\a\b}$. We
also write $N=(\l w)$ and $N^{ab}=(\l\g^{ab}w)$. The derivative
$w_\a={\*\over\*\l^\a}$ does not respect the pure spinor constraint
and is usually demanded to occur in such combinations. 

\section\MinBSec{The minimal $b$ operator}Let $V$ be the space of linear functions of an unconstrained spinor $\lambda^\alpha$.
Define $R=\Sym^\bullet(V)$ as the functions of $\lambda$. Functions of a pure spinor belong to the space
$S=R/I$, where $I$ is the ideal generated by $(\lambda\gamma^a\lambda)$.
The derivative $w_\alpha$ is an operator on $R$, but not on $S$.

There is, however, a generalisation of the derivative,
$$
\tw_\a=w_\a-\fr{4(N+3)}(\g^a\l)_\a(w\g_aw)\komma\Eqn\CovDer
$$
that is well defined
[\CederwallOperators,\AisakaArroyoBerkovitsNekrasov]. 
Its action between monomials $\l^{\a_1\ldots\a_n}$ in
the modules $(0000n)$ is identical to the action of $w$, but maps the
ideal generated by $(\l\g^a\l)$ to itself. 
This is indeed the most proper definition of $\tw$, since neither of the terms in eq. (\CovDer) is well defined on $S$. The concrete form is useful for explicit calculations, which then formally are performed in the space $R$, after which $I$ is modded out. This is consistent thanks to $(\tw I)/I=0$.
Note that double
contractions are needed for the gauge invariance (well-definedness) of
$\tw$, so it is a genuinely quantum mechanical operator, and that it
is not a ``covariant derivative''---it does not satisfy a Leibniz
rule. It follows from $\tw$ respecting the ideal that $(\tw\gamma^a\tw)=0$. This is also straightforwardly obtained from a direct calculation.
A useful relation in calculations is
$$
[\tw_\a,\l^\b]=\d_\a{}^\b-\fr{2(N+3)}(\g^a\l)_\a(\g_a\tw)^\b\punkt
\Eqn\WLCommutator
$$
It immediately leads to 
$$
[\tw_\a,\l^\a]=16-{5N\over N+3}\;,
$$
interpolating between $16$ (the dimension of the spinor module) at $N=0$ and $11$ (the dimension of pure spinor space) 
as $N\rightarrow\infty$.

Using $\tw$, it is possible to form operators of negative ghost number, built entirely out of minimal pure spinor variables.
One such operator is the $b$ operator.
In $D=11$ supergravity [\PureSGI,\PureSGII], or in any supersymmetric model where the constraint on $\lambda$ does not put it on a minimal orbit, demanding $\tw$ to respect the ideal does not uniquely define it [\CederwallOperators], and one will have access to more than one operator on $S$ with ghost number $-1$. The condition that it acts as $w$ between the modules in $S$ will specify it uniquely.

We will now show concretely how a minimal $b$ operator can be constructed, first by adding trivial terms to the non-minimal $b$, then by making a general Ansatz and solving it.

Normally, it is stated that $b_3$ represents $\db$ (operator)
cohomology. This may be
slightly confusing---since there is no 3-form cohomology, and any
operator cohomology should map between cohomologies, one might think
that $b_3$ is trivial. 
To understand this better, it is instructive to
examine exactly why $b_3$ is closed, and in what sense it is not
exact.

We use Dynkin labels for denoting highest weight modules, where $\lambda\in(00001)$, $\bar\lambda\in(00010)$.
The factor $\lb(d\lb)^3$ in $b_3$ comes in the module $(02000)$. The
prefactor $(\l\lb)^{-4}$ implies that $\{\db,b_3\}$ will contain 
$\lb(d\lb)^4$ antisymmetrised in the five spinor indices. There are
two compeletly antisymmetric modules, $(11001)$ and $(00003)$. The
second of these is not reached from $(00010)\otimes(02000)$, so one is
left with the first one. It is a $\g$-traceless 
3-index (hook) spinor, contracted with an object in the conjugate
module $(11010)$, formed from
one $\l$ and two $N_{ab}$'s, which thus takes the form containing a
leading term
$(\g^d\l)_\a N_{ab}N_{cd}$. However, since $N_{ab}$ is constructed
from $\l$, one has $(\g^b\l)_\a N_{ab}\propto(\g_a\l)_\a N$. There is
no room for the module $(11010)$, and $\{\db,b_3\}=0$. The closedness
of $b_3$ relies on $N_{ab}$ containing $\l$.

Why, on the other hand, is $b_3$ non-trivial? If 
``naked'' $w$'s where allowed, $b_3$ would be trivial and
formed as $[\db,a_2]$, where 
$a_2\propto(\l\lb)^{-3}(d\lb\g^{abc}d\lb)(\lb\g_aw)N_{bc}$. This is of
course not allowed since this operator is ill defined. But if we
replace $w$ by the well defined operator $\tw$ of eq. (\CovDer), we
have a well defined operator $a_2$ such that $b_3+[\db,a_2]=0$. It is
well defined on all holomorphic functions, \ie, on all elements of the
cohomology of $\db$, but not on arbitrary cochains, since it it is singular
on functions of a certain integer negative degree of homogeneity
($-2$) in
$\l$. This makes it likely that it is possible to continue the
procedure, and find $a=a_0+a_1+a_2$ such that $b'=b+[Q,a]$ only
contains a $0$-form term and is holomorphic, $\{q,b'\}=\square$ and
$\{\db,b'\}=0$. This can indeed be done, but at the price of obtaining
an operator which is singular on functions of some negative degree of homogeneity in $\l$.

At each step in the calculation (starting from $a_2$ and going down in
form degree) one needs $b_i+[q,a_i]+[\db,a_{i-1}]=0$. The
$\db$-exactness of the first two terms follows, roughly speaking, from
the extra factors $\l$ introduced by the commmutator (\WLCommutator). 
The operator $a$ is determined by a lengthy calculation to be
$$
\eqalign{
a&=a_0+a_1+a_2\komma\cr
a_0&=(\l\lb)^{-1}\Bigl[-\fr2\left(1-\Fr{N+6}{(N+4)(N+5)}\right)(\lb\g^a\tw)\*_a
-\fr{384(N+4)}(\lb\g^{abc}\tw)(D\g_{abc}D)\cr
&\qquad\qquad\qquad
+\fr{128(N+4)(N+5)}(\lb\g^a\tw)N^{bc}(D\g_{abc}D)\Bigr]\komma\cr
a_1&=\fr{384}(\l\lb)^{-2}(\lb\g^{abc}d\lb)
     \left[(\tw\g_{abc}D)-\Fr3{N+4}N_{ab}(\tw\g_cD)\right]\komma\cr
a_2&=\fr{128}(\l\lb)^{-3}(d\lb\g^{abc}d\lb)(\lb\g_a\tw)N_{bc}\punkt\cr
}\eqn
$$

The holomorphic $b'$ operator now takes the form
$$
b'=\fr{(N+4)(N+5)(N+6)}\left[-\fr2(N^2+9N+15)(\tw\g^aD)\*_a
+\fr{128}N^{ab}(\tw D^3_{ab})\right]\komma\Eqn\BPrime
$$
where $D^3_{ab}$ is the antisymmetric product of three $D$'s in
$(01001)$,
$$
\eqalign{
(D^3)_{ab}^\a&=(\g^i)^{\a[\b}(\g_{abi})^{\g\d]}D_\b D_\g D_\d\cr
&=(\g^i D)^\a(D\g_{abi}D)+\fr4(\g_{[a}\g^{ij}D)^\a(D\g_{b]ij}D)
      -\fr{72}(\g_{ab}\g^{ijk}D)^\a(D\g_{ijk}D)\cr
&=(\g^i D)^\a(D\g_{abi}D)+32(\g_{[a}D)^\a\*_{b]}
         -12(\g_{ab}{}^iD)^\a\*_i\punkt\cr
}\eqn
$$
Seen just as an operator on functions of $\l$, $x$ and $\theta$, 
it is not at all a
priori obvious that a $b'$ with $\{q,b'\}=\square$ should exist.
A general Ansatz would contain the two structures in eq. (\BPrime),
each multiplied by an arbitrary function of $N$:
$$
b'=f(N)(\tw\g^aD)\*_a+g(N)N^{ab}(\tw D^3_{ab})\punkt\eqn
$$
 The result of the
anticommutator $\{q,b'\}$ potentially contains three other structures
than $\square$, two with $D^2\*$ and one with $D^4$. The system
na\"\i vely looks overdetermined. The vanishing of
the $D^4$ term fixes the function $g(N)$ to be proportional 
to ${1\over(N+4)(N+5)(N+6)}$, while the
$\square$ term gives an recursion equation for the function $f(N)$,
$$
f(N)=-\Fr{N+3}{(N+6)(N+8)}\left(1+Nf(N-1)\right)\;.\Eqn\RecursionEq
$$
The general solution to eq. (\RecursionEq) is
$$
f(N)=-\Fr{N^2+9N+15}{2(N+4)(N+5)(N+6)}
+\Fr{c(-1)^N}{(N+1)(N+2)(N+3)(N+4)^2(N+5)^2(N+6)^2(N+7)(N+8)}\eqn
$$
(for some reason, there is a close resemblance of the homogeneous term
with the coefficients in the pure spinor Hilbert series, 
$$
Z(t)=\sum_{N=0}^\infty
\Fr{(N+1)(N+2)(N+3)^2(N+4)^2(N+5)^2(N+6)(N+7)}
{2\cdot3^2\cdot4^2\cdot5^2\cdot6\cdot7}t^N\punkt)\eqn
$$
Remarkably, when $c=0$. the
respective contributions of the two terms to each of the $D^2\*$ terms
in $\{q,b'\}$ take the same functional form in $N$, and the result is
$b'$ according to eq. (\BPrime).

Acting on a field $\psi=\l^\a A_\a$, the second term in eq. (\BPrime) gives $0$, and one immediately gets
$$
b'\psi=-\fr{16}\*^a(D\g_aA)\komma\eqn    
$$
which shows that $b'$ implies exactly Lorenz gauge.

\section\PhysOpSec{Physical operators}In ref. [\CederwallKarlssonBI], so called ``physical operators'' were defined, then with the purpose of writing possible higher-derivative corrections.
For $D=10$ super-Yang--Mills theory, these are ghost number $-1$ operators obeying a sequence of relations obtained from inspection of the pure spinor superfield equations of motion,
$$
\eqalign{
[Q,\hat A_\a]&=-D_\a-2(\g^i\l)_\a\hat A_i\komma\cr
\{Q,\hat A_a\}&=\*_a-(\l\g_a\hat\chi)\komma\cr
[Q,\hat\chi^\a]&=-\fr2(\g^{ij}\l)^\a\hat F_{ij}\komma\cr
\{Q,\hat F_{ab}\}&=\ldots
}\Eqn\Sequence
$$
with non-minimal solutions
$$
\eqalign{
\hat A_\a&=-(\l\lb)^{-1}\left[\fr8(\g^{ij}\lb)_\a N_{ij}
       +\fr4\lb_\a N\right]\komma\cr
\hat A_a&=-\fr4(\l\lb)^{-1}(\lb\g_aD)
       +\fr{32}(\l\lb)^{-2}(\lb\g_{a}{}^{ij}d\lb)N_{ij}\komma\cr
\hat\chi^\a&=\fr2(\l\lb)^{-1}(\g^i\lb)^\a\*_i
            -\fr{192}(\l\lb)^{-2}(\lb\g^{ijk}d\lb)(\g_{ijk}D)^\a\cr
&\qquad         -\fr{64}(\l\lb)^{-3}(\g_i\lb)^\a(d\lb\g^{ijk}d\lb)N_{jk}
        \komma\cr
\hat F_{ab}&=\fr8(\l\lb)^{-2}(\lb\g_{ab}{}^id\lb)\*_i
              +\fr{32}(\l\lb)^{-3}(d\lb\gamma_{ab}{}^id\lb)(\lb\g_iD)\cr
&\qquad          -\fr{256}(\l\lb)^{-4}(\lb\g_{abi}d\lb)(d\lb\g^{ijk}d\lb)N_{jk}
\komma\cr
\ldots&\cr
}\eqn
$$
Suppose we would like to find physical operators constructed from minimal variables.
We then observe that eq. (\WLCommutator) leads to 
$$
[q,\tw_\alpha]=-D_\alpha+\fr{2(N+3)}(\gamma^i\lambda)_\alpha(\tw\gamma_iD)\;.\eqn
$$
So, in the minimal picture we can identity 
$$
\eqalign{
\hat A_\alpha&=\tw_\alpha\;,\cr
\hat A_a&=-\fr{4(N+4)}(\tw\gamma_a D)\;.
}\eqn
$$
The rewriting of the $b$ operator in terms of physical operators in refs. [\BenShaharGuillen,\GuillenTamingElevenB]  is probably one of the deepest and physically most meaningful ones.
We expect it to hold equally in minimal variables.

%

\section\DiscSec{Discussion}We have seen how, contrary to what is sometimes stated in the literature, negative ghost number operators, such as the $b$ operator or physical operators, are quite naturally constructed in minimal pure spinor superspace. This will certainly apply also to the operators used in the action of $D=11$ supergravity [\PureSGI,\PureSGII] and to any other models
[\CederwallBLG\skipref\CederwallABJM\skipref\CederwallReformulation\skipref\CederwallDSix\skipref\CederwallExotic-\CederwallSLFive].
There may be a certain advantage in using the minimal operators. The interpretation of the operators typically is that they map between different cohomology classes, and those are most simply represented in the minimal picture. Therefore, using minimal operators yields again minimal representatives, without the need of adding trivial terms.

However, most likely,  $b'^2=\{q,\Omega\}$, $\Omega\neq0$. The first term in eq. (\BPrime) squares to zero, but we have not been able to show this for the rest.
This may present a drawback in comparison with the non-minimal $b$.

The algebraic behaviour of $b$ are essential for proving properties of amplitudes like color-kinematics duality and double copy
[\BenShaharGuillen,\BorstenKinematic]. Its failure to be a derivation defines the brackets for the kinetic Lie algebra. We have not examined this in detail for $b'$.

Finally, other versions of the pure spinor differential may present other possibilities to construct negative ghost number operators.
In ref. [\CederwallJonssonPalmkvistSaberi], instead of using constrained spinors, the Tate resolution of the constraint gives an alternative differential.

\vskip4\parskip
\noindent\underbar{Acknowledgements:} The author is grateful to Max Guillen and Leron Borsten for helpful discussions, and to Nordita, where part of this work was done, for hospitality.

\refout

\end